\documentclass[review,sort&compress]{elsarticle}

\usepackage{amsmath,bbm,color,gensymb,hyperref,multirow,multicol,setspace,ulem,soul}
\usepackage[usenames,dvipsnames,svgnames,table]{xcolor}
\usepackage[font=normal,labelfont=bf]{caption}
\usepackage[font=normal]{subcaption}
\usepackage[margin=1in]{geometry}
\usepackage{nomencl}
\makenomenclature
\hypersetup{colorlinks=true,linkcolor=blue,filecolor=magenta,urlcolor=cyan}
\journal{arXiv}

\begin{document}

\begin{frontmatter}

\title{Bayesian strategies for uncertainty quantification of the thermodynamic properties of materials}

\author[addr1]{Noah H. Paulson\corref{mycorrespondingauthor}}
\cortext[mycorrespondingauthor]{Corresponding author}
\ead{npaulson@anl.gov}

\author[addr2]{Elise Jennings}
\ead{ejennings@anl.gov}

\author[addr1]{Marius Stan}
\ead{mstan@anl.gov}

\address[addr1]{Applied Materials Division, Argonne National Laboratory, Lemont, IL 60439}
\address[addr2]{Leadership Computing Facility, Argonne National Laboratory, Lemont, IL 60439}

\begin{abstract}

Reliable models of the thermodynamic properties of materials are critical for industrially relevant applications that require a good understanding of equilibrium phase diagrams, thermal and chemical transport, and microstructure evolution. The goal of thermodynamic models is to capture data from both experimental and computational studies and then make reliable predictions when extrapolating to new regions of parameter space. These predictions will be impacted by artifacts present in real data sets such as outliers, systematics errors and unreliable or missing uncertainty bounds. Such issues increase the probability of the thermodynamic model producing erroneous predictions. We present a Bayesian framework for the selection, calibration and quantification of uncertainty of thermodynamic property models. The modular framework addresses numerous concerns regarding thermodynamic models including thermodynamic consistency, robustness to outliers and systematic errors by the use of hyperparameter weightings and robust Likelihood and Prior distribution choices. Furthermore, the framework's inherent transparency (e.g. our choice of probability functions and associated parameters) enables insights into the complex process of thermodynamic assessment. We introduce these concepts through examples where the true property model is known. In addition, we demonstrate the utility of the framework through the creation of a property model from a large set of experimental specific heat and enthalpy measurements of Hafnium metal from 0 to 4900K.

\end{abstract}


\end{frontmatter}

\newpage

\section{Introduction}
\label{sec:introduction}

High quality thermodynamic property models are the key to calculating numerous other properties and predicting complex material behaviors. Researchers construct these models through a process called assessment \cite{Kattner1997_MulticomponentPhaseEquilibria} whereby experimental and computational information regarding material properties is used to evaluate the optimal mathematical forms that relate properties to parameters such as temperature, pressure and composition.  Thermodynamic assessments present numerous challenges: the presence of outliers, missing or underestimated errors in datasets and systematic errors are commonplace. Furthermore, it is critical to select model forms that match the trends in the data and are based on the physics and chemistry of the material. These complex judgments are the sole responsibility of the practitioner and are rarely quantitatively expressed in the final models. Bayesian statistical methods provide an opportunity to address each of these challenges in a robust and comprehensive manner.

Over the past several decades, researchers have developed approaches for the quantification of uncertainty in thermodynamic property models. The challenges involved in the calculation of phase diagrams (CALPHAD) have driven many of these efforts. CALPHAD requires the selection and calibration of models for the Gibbs energies of relevant phases from disparate experimental and computational sources \cite{Kattner1997_MulticomponentPhaseEquilibria}. Through these procedures, researchers fit empirical model forms to limited and often conflicting datasets. Both frequentist \cite{Jansson1984_CalphadBook,Malakhov1997_PhaseBoundaryFrequentistUQ} and Bayesian \cite{Königsberger1991_CalphadSequentialBayes,Olbricht1994_UnaryBayes,Chatterjee1994_GeologyBayesPD,Chatterjee1998_ThermoDatabaseBayesian,Stan2003_GibbsBayesian,Duong2016_BayesianCalphad,Otis2017_CalphadBayesian} approaches have been employed to construct models that best represent the data, ensure thermodynamic consistency and provide estimates of parameter and model uncertainties. Furthermore, authors have employed such approaches to build thermodynamic property models for single-component systems \cite{Olbricht1994_UnaryBayes,Roslyakova2016_GibbsSegmentedRegression}.

While the previously described efforts made significant progress in evaluating uncertainty of thermodynamic properties, a general framework for model selection, calibration and uncertainty quantification is lacking. Bayesian statistical methods have the potential to address each of these elements in an elegant and self-consistent manner. Researchers in a number of fields have employed comprehensive Bayesian modeling approaches to great effect, with particular success in the cosmology community \cite{Verde2010_CosmologyStats,Ma2014_Hyperparameter}. In this work, we present a Bayesian framework for the construction of thermodynamic property models, with applications in the greater materials science community. This framework is enabled by recent advances in numerical sampling algorithms in Statistics which are available in several open source packages \cite{Forman2013_emcee,Feroz2013_MultiNestImportanceSampling,Farr2015_kombine}.

This Bayesian framework naturally accommodates ancillary methods to address issues commonly seen in real computational and experimental data, including the presence of outliers, systematic errors and inaccurate error bars. Our novel approach optimally leverages specific heat and enthalpy data (as well as entropies and Gibbs free energies if available) to construct self-consistent thermodynamic models. One major advantage of this approach versus previous attempts is that it does not incur the errors associated with the conversion of all data to the same property basis (e.g. fitting and then differentiating enthalpy data to convert to specific heat). Furthermore, the framework can naturally incorporate the relative weighting of various datasets and provides significant insight into the relative importance of each dataset to the model.

We demonstrate the strengths of the framework through the construction of a thermodynamic property model for Hafnium metal from a diverse collection of experimental measurements of heat capacity and enthalpy for the alpha, beta and liquid phases. Hafnium metal is a good candidate for this study as the available measurements exhibit the previously mentioned issues due to the difficulty of obtaining Hafnium samples without Zirconium content, the potential for oxygen contamination, and the extreme temperatures at which the beta and liquid phases are stable (Hafnium melts at $\sim$2500K and vaporizes at 4900K) \cite{Arblaster2014_HfAssessment}. In developing the thermodynamic property models we investigate model forms that are effective across the entire temperature range. In the low temperature regime we utilize the Einstein and Debye models of specific heat to account for quantum effects. At higher temperatures, both polynomial models and the segmented regression model developed by Roslyakova et al. \cite{Roslyakova2016_GibbsSegmentedRegression} are investigated. The resulting property models are compared with assessments performed using the HSC Chemistry software package \cite{Roine2002_HSC_Chemistry} and the recent assessment by Arblaster \cite{Arblaster2014_HfAssessment}. The comparison with these assessments provide an opportunity to compare the strategy of the Bayesian framework with that of a commercial software package and a human expert.

\section{The Bayesian Framework}
\label{sec:bayesian_framework}
In Section \ref{subsec:bayes_thm} we introduce Bayes' Theorem and statistics. In Section \ref{subsec:numerical_sampling} we describe some modern statistical sampling techniques which will be used in this paper.
In Section \ref{subsec:robustness_to_outliers} we discuss the model's robustness to outliers.
In Section \ref{subsec:data_quality} we describe the weighting applied to different datasets and in Section \ref{subsec:thermodynamic_consistency} we outline how thermodynamic consistency is enforced in the model.
We provide example problems to highlight the application of each method.
In Appendix A, we present a simple example of fitting a linear model to noisy data. 

\subsection{Bayes' Theorem}
\label{subsec:bayes_thm}

Bayes' theorem states that given some data, D, and a model, M, the probability density function of the model parameters, $\bf{\Theta}$, called the Posterior distribution, is given by,
\begin{equation} \label{eqn:bayes}
P\left(\boldsymbol{\Theta}| \boldsymbol{D}, M \right) = \frac{P\left(\bf{D}|\boldsymbol{\Theta},M\right) P\left(\boldsymbol{\Theta}| M\right)}{P\left(\boldsymbol{D}| M\right)},
\end{equation}
where $P\left(\bf{D}|\boldsymbol{\Theta},M\right)$ is the Likelihood which describes the conditional probability of the data given the model and associated parameters. $P\left(\boldsymbol{\Theta}| M\right)$ is the Prior and describes our previous beliefs or knowledge about the model parameters (e.g. ranges from previous experiments or physical restrictions on the parameter values). $P\left(\boldsymbol{D}| M\right)$ is the marginal Likelihood or Evidence and describes the probability of the data being generated by the model. A characteristic of Bayesian statistics that differentiates it from the frequentist approach is that both the parameters and the data are assumed to be random variables and we can make probabilistic statements of certainty given a single realization of the data. This has important implications in the interpretation of Bayesian probabilities.

The quantities which we want to evaluate are moments of the probability function (e.g. means, variances and higher order moments) which require integrating over the Posterior probability. For many problems this is difficult to evaluate and so the solutions are obtained using numerical methods such as Markov Chain Monte Carlo (MCMC) \cite{Gelman2013_BayesianBook} which seeks an approximate solution to the exact Posterior; or variational inference methods which find an exact solution to an approximation of the Posterior \cite{Blei2017_VariationalInferenceReview}.

An accurate determination of the Posterior probability enables us to measure their moments as well as uncertainty intervals (or credible intervals in the Bayesian vocabulary) for the model response. One advantage of the Bayesian framework is that uncertainties have intuitive explanations; for example, a one standard-deviation interval for a parameter of interest contains  $\sim$68\% of its probability distribution (assuming Gaussian distributed variables). In the remainder of this work, the 2.5$^{\mbox{th}}$ and 97.5$^{\mbox{th}}$ percentile bounds form the 95$^{\mbox{th}}$ percentile uncertainty intervals, and the best fit parameters and model predictions are given by the 50$^{\mbox{th}}$ percentile levels of the quantities of interest.

Bayesian statistics also provides a strategy for model selection via the marginal Likelihood. The marginal Likelihood is the denominator of Eq. (\ref{eqn:bayes}) and represents the probability of the data given the model. The marginal Likelihood has the desirable qualities of rewarding models that match the data well and penalizing models that are overly complex (i.e. have too many degrees of freedom or parameters). The marginal Likelihood is given by,
\begin{equation} \label{eqn:marginal_likelihood}
P\left(\boldsymbol{D}| M\right)=\int_{\Omega_{\boldsymbol{\Theta}}}P\left(\bf{D}|\boldsymbol{\Theta},M\right)P\left(\boldsymbol{\Theta}| M\right)d\boldsymbol{\Theta},
\end{equation}
where $\Omega_{\boldsymbol{\Theta}}$ represents the complete parameter space. Note that by integrating over the parameters in the model we are able to compare Bayesian Evidences for two distinct models in a meaningful way, even if they contain different numbers of parameters. This ratio of the Bayesian Evidence for two models is called the Bayes Factor and is given by
\begin{equation} \label{eqn:bayes_factor}
R=\frac{P\left(M_A| \boldsymbol{D}\right)}{P\left(M_B| \boldsymbol{D}\right)}=\frac{P\left(\boldsymbol{D}| M_A\right)P\left(M_A\right)}{P\left(\boldsymbol{D}| M_B\right)P\left(M_B\right)},
\end{equation}
where $M_A$ and $M_B$ are the two models under consideration \cite{Feroz2013_MultiNestImportanceSampling}. Kass and Raftery \cite{Kass1995_BayesFactor} provide a commonly used guide to interpret the Bayes factor, in which a value in the range $3-20$ indicates a positive strength of evidence for the preference of one model versus another and a factor in the range $20-150$ represents strong evidence to prefer one model over another. Unfortunately because of the integral over parameter space Eq. (\ref{eqn:marginal_likelihood}) is notoriously difficult to evaluate. Only with recent advances in sampling techniques has the model Evidence become an appealing option for model selection. We briefly discuss some selected sampling approaches in the following section. 

\subsection{Numerical Sampling}
\label{subsec:numerical_sampling}

Analytical methods to evaluate the Posterior probability are only available for a small subset of problems and in practice numerical sampling approaches are employed for this task. The most common and historically important sampler is the Metropolis-Hastings (M-H) algorithm \cite{Chib1995_M-H_Algorithm}, which makes successive jumps in parameter space (based on a proposal distribution), that are accepted when the ratio of the current to the previous local Posterior probability exceeds a uniformly distributed random number between zero and one. While the M-H algorithm is guaranteed to obtain the Posterior distribution after a sufficient number of steps, it is inefficient when applied to non-trivial distributions. For example, it is a significant challenge to determine a proposal distribution that does not lead to a precipitous drop in sampling efficiency for distributions with degeneracy (the Rosenbrock function is a classic example). Furthermore, the M-H algorithm struggles to sample multimodal distributions, as the hops must traverse regions of low Posterior probability (and therefore low acceptance probability).

In this work, we discuss two state-of-the-art samplers that improve on the M-H algorithm: \textit{kombine} \cite{Farr2015_kombine} and \textit{MultiNest} \cite{Feroz2013_MultiNestImportanceSampling}. kombine is based on Goodman and Weare's affine-invariant ensemble sampler \cite{Goodman2010_AffineInvariantEnsembleSampler}. Goodman and Weare's algorithm leverages the positions of a set of `walkers' (i.e. independently initialized samplers) to propose a jump with a high probability of acceptance (even for degenerate distributions). kombine  builds on this by defining local estimates of the Posterior probability function defined by $K$ nearest neighbor clusters of walkers which update a local Kernel Density Estimate (KDE) of the distribution \cite{Silverman2018_DensityEstimationBook}. This estimate enables both frequent jumps between modes and the computation of the marginal Likelihood. It is important to note that the researcher must monitor MCMC-based methods such as kombine to ensure that the walkers effectively traverse the Posterior distribution and that subsequent jumps are not overly correlated \cite{Gelman2013_BayesianBook}. Many approaches exist to monitor convergence. The Gelman-Rubin statistic compares within-sequence and between-sequence variance to assess how well the walkers explore the Posterior (mixing) and whether each walker has achieved convergence (stationarity) \cite{Gelman2013_BayesianBook}. The Gelman-Rubin statistic should be close to one, with 1.1 being a conservative limit. The Gelman-Rubin statistic is discussed further in Appendix A. Additionally, the effective number of independent simulation draws may be employed to ensure that a statistically significant number of independent samples have been drawn from the posterior \cite{Gelman2013_BayesianBook}. Lastly, the researcher must visually inspect the sequences to assess mixing and stationarity.

In contrast to other Monte Carlo based approaches, nested sampling \cite{Skilling2004_NestedSampling} does not utilize the concept of the Markov chain. The essential idea of this approach is to transform the multivariate integral for the marginal Likelihood (Eq. (\ref{eqn:marginal_likelihood})) into a one dimensional integral over regions of the Prior distribution bounded by iso-likelihood contours. In practice, a set of live points are sampled from the Prior and ranked by their respective Likelihoods. Each iteration, the lowest Likelihood point is removed from the set of live points, a new point is sampled from the Prior and this point is added to the set if it has a greater Likelihood than that of the point just removed. With each iteration, the live points occupy a smaller volume of the Prior until the iso-likelihood contour encloses an acceptably small volume with an associated error in the marginal Likelihood. This process efficiently computes the marginal Likelihood, and the resulting set of all current and previous live points are samples from the Posterior. MultiNest improves upon basic nested sampling by preferentially sampling regions of the Prior more likely to fall within the current iso-likelihood contour. In contrast to the M-H and kombine algorithms, MultiNest requires little supervision, proceeding until the estimated marginal Likelihood reaches a maximum within some pre-specified tolerance \cite{Feroz2013_MultiNestImportanceSampling}. In this work, we employ the open source python implementation of the MultiNest algorithm, \textit{pymultinest} \cite{Buchner2014_pymultinest}.

\subsection{Robustness to Outliers}
\label{subsec:robustness_to_outliers}

It is common practice in the development of thermodynamic property models to identify outliers and remove them from the analysis. This may be a reasonable approach in many cases, especially when the questionable data points or sets have known sources of error or capture different physical phenomena. In other cases, however, we may not know the reasons for the discrepancy, and we risk ignoring relevant physics or uncertainty in the property of interest by removing the data. In the following section, we present an approach to incorporate datasets with potential systematic errors.

In the previous example, we assume a Normal distribution for the Likelihood. While this is a reasonable assumption when each datum represents the mean of multiple data points, it does not necessarily hold in all cases. The use of a distribution with higher probability density in the tails, such as Student's-t distribution have a better chance of robustly accommodating outliers. The probability density function of Student's-t distribution is given by
\begin{equation} \label{eqn:students-t}
f\left(t\right)=\frac{\Gamma\left(\frac{\nu+1}{2}\right)}{\sqrt{\nu\pi}\Gamma\left(\frac{\nu}{2}\right)}{\left(1+\frac{t^2}{\nu}\right)}^{-\frac{\nu+1}{2}},
\end{equation}
where $\nu$ is the degrees of freedom and $\Gamma$ is the gamma function. Gelman et al. suggest that it is appropriate to either select a reasonable value for the degrees of freedom, $\nu$, or to include it as a parameter in the Bayesian analysis if finding the specific form of the error's distribution is important \cite{Gelman2013_BayesianBook}. They also caution against employing $\nu <= 2$, as this $\nu$ results in a distribution with infinite variance that is not realistic in the far tails. One significant advantage of the Bayesian framework is that modifying the choice for the Likelihood is, in this case, trivial to implement and does not add to the computational overhead when evaluating the Posterior using numerical samplers. We show an example of this in Appendix B.

\subsection{Data Quality}
\label{subsec:data_quality}

In this section we introduce a technique in Bayesian inference when multiple datasets of varying quality are available. Several problems often arise in this situation. Firstly, journal articles may report errors for individual datasets, but these are not always reliable and must be independently evaluated. Secondly, some datasets may exhibit systematic errors where their mean behavior significantly deviates from the ‘true’ trend. In a method introduced by Lahav et al. \cite{Lahav2000_hyperparameter}, and expanded by Ma et al. \cite{Ma2014_Hyperparameter}, the reported errors for each dataset are weighted using hyperparameters in the Bayesian analysis. In the thermodynamics literature, data variances are often available, though covariances are rarely reported. Consequently, we do not present methods for handling data covariances in this work and we assume independence of the data points.

The hyperparameters for each dataset are incorporated into the Likelihood function as follows. Consider we have N datasets, each with t data points, we write the Likelihood as
\begin{equation} \label{eqn:ex3_likelihood}
P\left(\bf{D}| \bf{\Theta}, \bf{\alpha}, M\right) = \prod_i^N \prod_j^t \mathcal{N}\left(y_j^i| M\left(x_j^i,\bf{\Theta}\right),\varepsilon_j^i/\alpha^i\right),
\end{equation}
where $\varepsilon_j^i$ is the reported error (expressed as 1 standard deviation) of data point $x_j^i$ from dataset $D_i$ and $\alpha^i$ is the hyperparameter for dataset $D_i$. $\mathcal{N}\left(x|a,b\right)$ Denotes the probability density of a $x$ for a Normal distribution with mean $a$ and variance $b$. Note that while the reported errors for points within a data set may vary, we assign a single hyperparameter for each dataset. A small value for the hyperparameter increases the effective errors of the dataset and implies that the reported errors underestimate the true errors, while the inverse is true when the hyperparameter is large. Furthermore, a value of unity for the hyperparameter implies that the reported errors are accurate and that no rescaling is required. An alternate interpretation of the hyperparameter is as a measure of the relative importance of each dataset, with a low value indicating that the associated dataset has been de-emphasized in the analysis as compared to its original weighting. We define an Exponential Prior for each hyperparameter. The principal reason for this selection is that the Exponential distribution, with a rate of unity, has a mean of one, which is consistent with the initial expectation that the errors for each dataset have been properly reported. We present an example problem to demonstrate the advantages of this approach in Appendix C.

\subsection{Thermodynamic Consistency}
\label{subsec:thermodynamic_consistency}

The basic thermodynamic properties of materials, including specific heats, enthalpies, entropies and Gibbs free energies are of great utility to the scientific community. These quantities are connected by thermodynamic relationships that must be considered in the construction of thermodynamic property models. For example, the specific heat of a material is the derivative of the enthalpy with temperature. Material property databases generally report thermodynamic property models in the form of polynomial relationships for room temperature and above, in mathematical forms that include a common set of parameters \cite{Dinsdale1991_SGTEdata}. These parameters are typically obtained through a regression of experimental or computational information for a specific quantity. Sometimes all available data sets are leveraged by converting them to a common quantity. Conversions between quantities may be performed by direct differentiation and integration of the raw quantities or by first fitting models to the data followed by differentiation or integration. Unfortunately, such conversions are subject to unavoidable errors.

In this work, we propose a novel method to simultaneously utilize all available thermodynamic information (without conversion) to construct a common thermodynamic property model family for all quantities. Specifically, we express the overall Likelihood as the product of Likelihoods for each original quantity (and associated property model) and define priors for the parameters. By specifying a single prior distribution for each parameter common to multiple property models, we ensure that the final model family will be thermodynamically consistent and optimally leverage diverse thermodynamic information. We illustrate these concepts in the remainder of this section using synthetic datasets of specific heat and enthalpy.

Firstly, we generate 100 synthetic data points each for specific heat and enthalpy. The specific heats and enthalpies are equally spaced in the $1K-75K$ and $300K-1800K$ temperature ranges, respectively. We select these non-overlapping regions of measurement for specific heat and enthalpy to demonstrate that the Bayesian framework can capture robust representations of both quantities when data does not cover the entire temperature range for each quantity. We generate synthetic data for the enthalpy, $H$ and specific heat, $C_p$, via Eq. (\ref{eqn:ex4_enthalpy}) and Eq. (\ref{eqn:ex4_specific_heat}), respectively,
\begin{subequations} \label{eqn:ex4_models}
\begin{gather}
H\left(T\right)-H\left(298.15K\right)=\frac{3R\theta}{e^{^\theta/_T}-1}+a\frac{T^2}{2}+b\frac{T^3}{3}   \label{eqn:ex4_enthalpy}   \\
C_p\left(T\right)=\frac{d}{dT}H\left(T\right)=3R{\left(^\theta/_T\right)}^2\frac{e^{^\theta/_T}}{{e^{^\theta/_T}-1}^2}+aT+bT^2,   \label{eqn:ex4_specific_heat} 
\end{gather}
\end{subequations}
where R is the gas constant and the true model parameters are defined as follows: $\theta=150 K$, $a=6 \times 10^{-3} J/ mol. K$, $b=7 \times 10^{-7} J/ mol. K^2$ and $c=-5 \times 10^3 J/ mol.$ We add Gaussian noise with a standard error of $5000 J/ mol.$ to the enthalpy values. We add Gaussian noise to the specific heat values that scales linearly with the magnitude to a maximum of $3 J/ mol. K$ at $75 K$. Note that the model form of the specific heat in Eq. (\ref{eqn:ex4_specific_heat}) was proposed at the Ringberg workshop in 1995 to be effective across the entire temperature range \cite{Chase1995_RingbergWorkshop}. In this case, the first term employs the Einstein model to account for low-temperature behavior, though the Debye or other models may be substituted \cite{Grimvall1999_ThermophysPropMaterials}.

As mentioned previously, we define our Likelihood as the product of the individual Gaussian Likelihoods for the enthalpy and specific heat, i.e.,
\begin{equation} \label{eqn:ex4_likelihood}
P\left(\boldsymbol{D}; \boldsymbol{\Theta}, M\right)=\prod_{i=1}^{100}\mathcal{N}\left(D_{H,i}| H\left(T_i,\boldsymbol{\Theta}\right),\varepsilon_i\right)\prod_{i=1}^{100}\mathcal{N}\left(D_{C_p,i}| C_p\left(T_i,\boldsymbol{\Theta}\right),\varepsilon_i \right),
\end{equation}
where $\boldsymbol{D}$ represents the data, $\boldsymbol{\Theta}$ is the set of parameters, $\{\theta, a, b\}$, $D_{H,i}$ are enthalpy data and $D_{C_p, i}$ are specific heat data. At first, we initialize priors conservatively to cover the entire parameter ranges. We then iteratively narrow the priors to include the bulk and tails of the posterior distribution. The final priors for the four parameters are given as follows,
\begin{subequations} \label{eqn:ex4_priors}
\begin{gather}
\theta \sim \mathcal{U}\left(\theta| 145,155\right)    \label{eqn:ex4_prior_theta}   \\
a \sim \mathcal{U}\left(a| 0.003,0.009\right)    \label{eqn:ex4_prior_a}    \\
b \sim \mathcal{U}\left(b|-5\times 10^{-6},5\times 10^{-6}\right).    \label{eqn:ex4_prior_b}
\end{gather}
\end{subequations}

The Posterior probability is then evaluated using the MultiNest sampler with 800 live points. We carry out two different analyses. Firstly, we perform separated analyses where the models for specific heat and enthalpy do not share parameter values. Both Posterior distributions are simultaneously evaluated using MultiNest to obtain the combined marginal Likelihood for enthalpy and specific heat. The Likelihood expression for this analysis is identical to Eq. (\ref{eqn:ex4_likelihood}), save that the parameters are not shared between the models for enthalpy and specific heat (e.g., $\theta_{C_p}$ and $\theta_H$ are distinct parameters). Secondly, we perform a combined analysis using the Likelihood given in Eq. (\ref{eqn:ex4_likelihood}) where the model parameters are shared between the enthalpy and specific heat models. In other words, the analysis only employs a single parameter each for $\theta$, $a$ and $b$. Note that we select priors for the separate analysis in the same manner as in the combined analysis.

\begin{figure}
\centering
\includegraphics[width=1.0\linewidth]{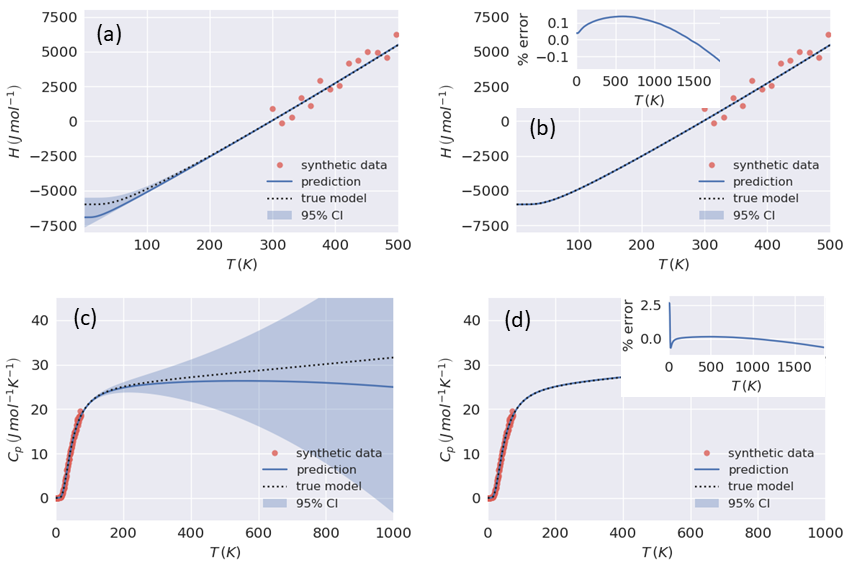}
\caption{Specific heat data, predictions and uncertainty intervals for enthalpy (a) using only enthalpy data, and (b) through a combined analysis; and for specific heat (c) using only specific heat data, and (d) through a combined analysis. Note that in panels b (d) the inset figure shows the ratio of the analyses using only enthalpy (specific heat) data  to the combined analysis  as a \% difference.}
\label{fig:ex4_pred}
\end{figure}

Figure \ref{fig:ex4_pred} presents the model predictions and uncertainty intervals for both the ‘separate’ and ‘combined’ analyses. For both quantities, the prediction from the ‘combined’ analysis nearly exactly reproduces the true model and exhibits tight uncertainty intervals. Additionally, the predictive accuracy extends well beyond the temperature ranges where data for the original quantity are available. As expected, the predictions from the ‘separate’ analysis exhibit large uncertainty intervals and poor model predictions due to the lack of data across the temperature range.

\begin{figure}
\centering
\includegraphics[width=.8\linewidth]{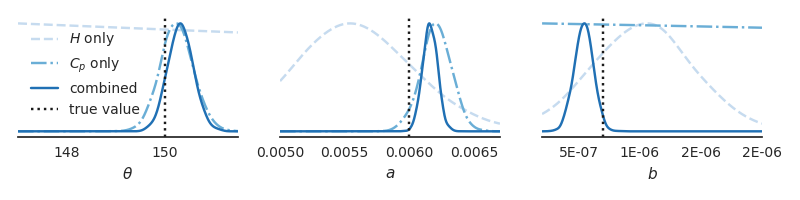}
\caption{1D Marginal Posterior distributions from both combined and separate $H$ and $C_p$ analyses. The distributions are vertically rescaled for ease of comparison. True parameter values are indicated by the vertical dotted black lines.}
\label{fig:ex4_1d}
\end{figure}

We find it informative to visually inspect the posterior distributions for both analyses. Figure \ref{fig:ex4_1d} displays univariate Posterior distributions alongside the true parameter values for each parameter. The best fit parameter values for the analysis where only enthalpy data are employed are poor for $\theta$ and $a$, and the reverse is true for the specific heat only analysis. In comparison, the combined analysis results in reasonable best fit values for all parameters and lower variance than for the separate analysis. Lastly, the Bayes’ Factor between the marginal Likelihoods for the two analyses was 29, corresponding to a strong preference for the combined approach presented here. In part, this preference likely stems from the reduced number of parameters required.

\section{Case Study}
\label{sec:case_study}

In this section we employ the Bayesian framework and methods introduced in the previous sections for the development of a thermodynamic property model for the alpha, beta and liquid phases of Hafnium.
In Section \ref{subsec:hafnium} we give an overview of Hafnium. In Section \ref{subsec:dataset_collection_and_correction} we discuss how the data was collected and any corrections which were applied. In Section \ref{subsec:analysis_methodology} we present our analysis methods and in Section \ref{subsec:model_selection} the model selection criteria are presented. Our main results are given in Section \ref{subsec:results}.

\subsection{Hafnium transition metal}
\label{subsec:hafnium}

Hafnium (Hf) is a tetravalent transition metal, commonly found in zirconium minerals. It is a good absorber of neutrons, which makes it a good material for control rods in some nuclear energy applications. For a review of the development of hafnium and comparison with other pressurized water-reactor control rod materials, see \cite{Keller1982_HafniumControlRod}. Hf also has a very high melting point, leading to applications in plasma welding torches \cite{Wang2012_HafniumCarbideCoating}. Superalloys of Hafnium, niobium, tungsten and titanium have excellent mechanical properties \cite{Hou2010_HafniumSuperalloyCreep}. Hafnium zirconate is a good gate dielectric for materials in electronics \cite{Hegde2007_HafniumZirconate}. Hafnium oxides, nitrides and carbides have additional applications that are outside the scope of this paper. The construction of a thermodynamic property model for Hafnium is interesting due to its high melting ($\sim 2500K$) and vaporization temperatures ($\sim 4900K$). This means that many more low temperature investigations are available than high temperature ones. The remainder of this work explores the thermodynamic properties of Hafnium in detail, yielding a  comprehensive assessment and, to the best of our knowledge, the only known assessment that includes both parameter uncertainties and model Evidences. Furthermore, through the application of the methods described in Sec. \ref{sec:bayesian_framework}, we explore the relative weighting of various datasets using hyperparameters in a novel way.

\subsection{Dataset Collection and Correction}
\label{subsec:dataset_collection_and_correction}

We have collected all available specific heat and enthalpy data of Hafnium metal for the alpha, beta and liquid phases from the published texts, correcting for Zirconium content and temperature scale when appropriate\footnote{Note that the measured values were extracted from tables or from figures using the free online WebPlotDigitizer utility \cite{Rohatgi2011_WebPlotDigitizer}}. We only exclude the measurements of Cristescu \cite{Cristescu1934_Hf_Cp_lowT} as the work makes no mention of the purity of the Hafnium sample. We prioritize original measurements over the published fits of the measurements. When available, we also collect the reported measurement errors. If errors were not reported, we assume +/- 5\% absolute error bounds. This is a conservative, but not unreasonable estimate. We convert errors to standard uncertainty (+/- 1 standard deviation) according to the GUM guidelines \cite{JCGM2008_UncertaintyGUM}. We assume that absolute error bounds represented the minimum and maximum of a Uniform distribution for the purposes of converting to a Normal distribution and extracting standard uncertainties \cite{JCGM2008_UncertaintyGUM}.

We perform corrections to account for differences in temperature scale standards \cite{Goldberg1992_TemperatureScale} and Zirconium contamination \cite{Arblaster2014_HfAssessment}. As papers rarely report the temperature scale in use at the time of publication, we assume that the standard temperature scale during the period of publication was utilized. Consequently, we convert measurements from the original scales (i.e. IPTS-48 and IPTS-68) to the ITS-90 standard. The maximum temperature difference due to a change in scale is 0.1\%, so we do not expect this correction to have a significant impact on the final analysis. Due to the chemical similarities between Zirconium and Hafnium, it is infamously difficult to obtain Hafnium specimens of high purity. Many thermodynamic measurements were performed on specimens with up to 3.12\% Zirconium content. If the original report does not contain corrections for Zirconium, we correct the thermodynamic properties for Zirconium content using the Kopp-Neumann Rule \cite{Arblaster2014_HfAssessment} and the thermodynamic properties of Zirconium assessed by Arblaster \cite{Arblaster2013_ZirconiumAssessment}. The maximum change in specific heat or enthalpy due to Zirconium content is 1.7\% for any dataset.


\subsection{Analysis Methodology}
\label{subsec:analysis_methodology}

We construct thermodynamic property models for the three phases of Hafnium metal via the framework introduced in Sec. \ref{sec:bayesian_framework}. As demonstrated in Sec. \ref{subsec:thermodynamic_consistency}, we can use a Bayesian framework to analyze the specific heat and enthalpy measurements in concert to ensure thermodynamic consistency and to optimally leverage the raw experimental measurements. We employ a Student's-t distribution with $2+1 \times 10^{-6}$ degrees of freedom for the Likelihood to correctly account for outliers, and utilize the approach of Sec. \ref{subsec:data_quality} to optimally rescale the reported errors. We select Uniform prior distributions for the model parameters (except for the hyperparameters of Sec. \ref{subsec:data_quality} for which Exponential prior distributions are used) to simplify the process of restricting parameter values to physically realistic domains. We use the MultiNest algorithm to sample the Posterior distributions because it outperforms the kombine sampler on time to solution when the number of parameters included in the Bayesian analysis becomes large (23 parameters for the most complex model). We use 800 live points in the MultiNest analysis. We have verified that no significant changes occur in the marginal Likelihood or Posterior distribution mean and uncertainty intervals when the number of live points is increased to 1600\footnote{Note that we use a threaded version of MultiNest which utilizes  all 36 cores on a single E5-2695v4 processor on the Bebop cluster at Argonne National Laboratory. The majority of the analyses run to completion within 3 hours on the compute cluster.}.

The specific Prior distributions chosen can affect the final Posterior distribution and the marginal Likelihood. Consequently, we take measures to consistently refine the Prior distributions for each model to ensure reliable estimates of these quantities. For each model of interest, we first select wide prior distributions which are guaranteed to encompass the expected parameter ranges. We perform the Bayesian analysis to obtain the initial Posterior distribution (note that due to the wide Posteriors, the MultiNest algorithm tends to require additional time to reach a solution). Following this we refine the prior boundaries to encompass the Posterior mean and 5 standard deviations on either side (assuming a Normal distribution). When the value of a parameter is subject to some physical or mathematical constraint, the constraint is prioritized (for example, the Debye temperature must be greater than zero). In Sec. \ref{subsec:results}, we investigate the effect of this approach on the marginal Likelihood and Posterior distribution of the final model in more detail.

\subsection{Model Selection}
\label{subsec:model_selection}
A critical aspect of the assessment is the identification of a potential set of model formulations. Model identification for the alpha phase requires the most care as the models have to be reasonable at cryogenic temperatures where quantum effects dominate. In this work, low temperature physics is captured by the Einstein and Debye formulations,
\begin{subequations} \label{eqn:cs_lowT}
\begin{gather}
C_p^{Ein}\left(T,\theta_E\right) = \frac{3R\theta_E}{e^{^{\theta_E}/_T}-1}    \label{eqn:cs_ein}   \\
C_p^{Deb}\left(T,\theta_D\right) = 9R{\left(\frac{T}{\theta_D}\right)}^3\int_0^{\theta_D/T}\frac{x^4e^x}{{\left(e^x-1\right)}^2}dx,   \label{eqn:cs_deb}
\end{gather}
\end{subequations}
where $\theta_E$ and $\theta_D$ are the Einstein and Debye temperatures, respectively. Unfortunately, Eq. (\ref{eqn:cs_deb}) cannot be evaluated analytically, so in this work we employ Simpsons integration with 100 equally-spaced points to evaluate the integral. Polynomial expressions (not including the constant term) are added to the low-temperature models to capture high-temperature properties \cite{Chase1995_RingbergWorkshop}. Additionally, we employ a segmented regression model developed by Roslyakova et al. \cite{Roslyakova2016_GibbsSegmentedRegression}. In this four parameter model, two linear segments are smoothly connected by a quadratic bend. The bend is parameterized by a location and bend half-width. In principle, this form might both model electronic effects at low temperatures and high temperature behavior. We employ the family of increasing order polynomials given by Eq. (\ref{eqn:ex1_model}) to model the higher temperature beta and liquid phases where quantum effects are not important. The development of thermodynamic property models that are effective across the entire temperature range is still an active area of research, so by no means is our model set exhaustive. Also, note that no constant term is employed in the enthalpy relationship, as we force the model to give zero enthalpy at 298.15 K (as is standard in the thermodynamic literature for the enthalpy of elements). The complete set of models evaluated in this work is given in Table \ref{tab:model_selection}.

Given a set of appropriate thermodynamic property models for specific heat and enthalpy for each phase, we apply the Bayesian framework to compute the entropy and Gibbs free energy for all phases. Due to the difficulty of analytically integrating the Debye relationship, the change in entropy in each phase relative to the lower transformation temperature, $T_{trans}$, was numerically computed according to the following relationship:
\begin{equation} \label{eqn:cs_s_rel}
S\left(T\right)=\int_{T_{trans}}^T\frac{C_p\left(\tilde{T}\right)}{\tilde{T}}d\tilde{T}.
\end{equation}

The entropy associated with each phase transformation is simply given by
\begin{equation} \label{eqn:cs_s_trans}
S_{trans}=\frac{H_{trans}}{T_{trans}},
\end{equation}
where $H_{trans}$ is the enthalpy of transformation. The entropy at any temperature is given by the entropy relative to the lower phase transformation temperature (e.g. $T_{\alpha\rightarrow\beta}$ for the beta-phase) plus the lower temperature cumulative relative entropies and entropies of transformation. The Gibbs free energy is computed according to the following relationship:
\begin{equation} \label{eqn:cs_g}
G\left(T\right)=H\left(T\right)-T\cdot S\left(T\right).
\end{equation}

\subsection{Results}
\label{subsec:results}

\begin{table}
\centering
\caption{Potential models, their log Marginal-Likelihoods and Bayes' Factors with respect to the selected model (marked with an asterisk) are presented for each phase.} 
\label{tab:model_selection} 
\begin{tabular}{cccc}
Model & Number of Model Parameters & Log Marginal Likelihood & Bayes' Factor\\
$\boldsymbol{\alpha}$\textbf{-phase} & & &\\
\hline
Einstein & 1 & -1744.2 & $\sim 0$\\
Debye & 1 & -1262.9	& $\sim 0$\\
Debye + Linear & 2 & -1072.6 & $\sim 0$\\
Debye + Quadratic & 3 & -813.2 & $\sim 0$\\
Debye + Cubic & 4 &	-640.2 & $\sim 0$\\
Debye + Quartic* & 5 & -623.1 & $1$\\
Debye + Quintic	& 6 & -627.4 & $1.4\times 10^{-2}$\\
Debye + SR & 5 & -629.7 & $1.4\times 10^{-3}$\\
$\boldsymbol{\beta}$\textbf{-phase} & & &\\
\hline
Constant & 2 & -534.2 & $\sim 0$\\
Linear & 3 & -511.1 & $3.0\times 10^{-3}$\\
Quadratic* & 4 & -505.3 & $1$\\
Cubic & 5 & -518.5 & $1.9\times 10^{-6}$\\
\textbf{Liquid-Phase} & & &\\
\hline
Constant & 2 & -491.4 & $\sim 0$\\
Linear* & 3 & -471.0 & $1$\\
Quadratic & 4 & -476.0 & $6.7\times 10^{-3}$\\
\end{tabular}
\end{table}
Table \ref{tab:model_selection} summarizes the potential models for the alpha, beta and liquid phases. Model names refer to the form of the specific heat relationship. For example, ‘Debye + Quadratic’ refers to a specific heat model where the Debye relationship accounts for low temperature effects and a quadratic polynomial accounts for high temperatures. As previously mentioned, the constant term was excluded for the alpha phase models because the specific heat must be 0 at 0K. Alongside the model names, the table lists total number of parameters, marginal Likelihoods and the Bayes’ Factors. Note that for each phase the Bayes’ Factor for that model is the ratio of the marginal Likelihood for the model and the model marked with an asterisk. We only evaluate polynomial models up to the order at which the marginal Likelihood stops increasing. We select the ‘Debye + Quartic’, ‘Quadratic’ and ‘Linear’ models for the alpha, beta and liquid phases, respectively, as each exhibits at least a strong evidence of preference versus the next best model according to the Bayes’ Factors. Equations (\ref{eqn:Cp_alpha}) through (\ref{eqn:H_liquid}) present the selected expressions for the specific heat and enthalpy for all phases. 
\begin{subequations} \label{eqn:final_expressions}
\begin{gather}
C_p^{\alpha}\left(T\right) = C_p^{Deb}\left(T,\theta_D\right) + a_2T + a_3T^2 + a_4T^3 + a_5T^4     \label{eqn:Cp_alpha}   \\
H^{\alpha}\left(T\right)-H^{\alpha}\left(298.15K\right)=\int_0^T C_p^{Deb}\left(\tilde{T},\theta_D\right)d\tilde{T} + a_2\frac{T^2}{2} + a_3\frac{T^3}{3} + a_4\frac{T^4}{4} + a_5\frac{T^5}{5}    \label{eqn:H_alpha}    \\
C_p^{\beta}\left(T\right)=b_1+b_2T+b_3T^2    \label{eqn:Cp_beta}    \\
H^{\beta}\left(T\right)=b_0+b_1T+b_2\frac{T^2}{2}+b_3\frac{T^3}{3}    \label{eqn:H_beta}    \\
C_p^{liq}\left(T\right)=c_1+c_2T    \label{eqn:Cp_liquid}    \\
H^{liq}\left(T\right)=c_0+c_1T+c_2\frac{T^2}{2}    \label{eqn:H_liquid}
\end{gather}
\end{subequations}

\begin{figure}
\centering
\includegraphics[width=0.6\linewidth]{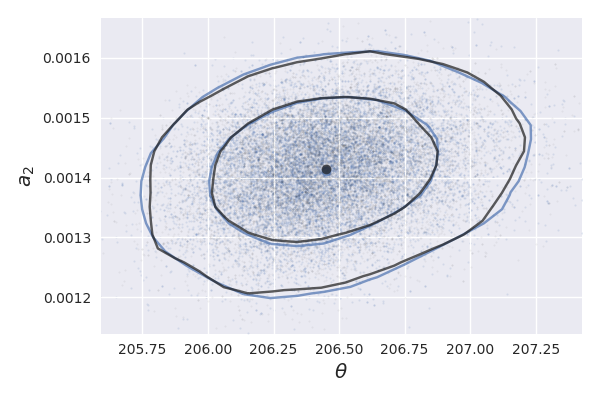}
\caption{The 68\% (95\%) uncertainty intervals are shown as the inner (outer) contours.  Results using unrestricted priors are shown as blue lines and restricted priors as black lines for both contours. The dots represent samples from the Posterior distribution and the circular markers represent the means of the Posterior distribution for both Prior choices (Note that these lie on top of one another in this figure).}
\label{fig:final_prior_selection}
\end{figure}

As mentioned previously, we carry out two stages of analysis; firstly using broad unrestrictive priors to identify high probability regions of parameter space (stage A), followed by a subsequent analysis with tighter priors (stage B). We demonstrate that this does not impact model selection, as we find that the rank-ordering of the models is identical before and after tightening the priors. As an example, we examine the effect of narrowing the priors for the selected alpha-phase model. Changing the priors increases the log marginal Likelihood from -641.1 to -623.1. In contrast, the Posterior distributions change little between the stages A and B, with the maximum relative errors in mean and standard deviation for any parameter in the Posterior distribution limited by 1\% and 5\%, respectively. Figure \ref{fig:final_prior_selection} illustrates the effect of changing the Prior on the Posterior distribution of the $\theta$ and $a_2$ parameters. The original samples, means and uncertainty intervals (68\% and 95\%) are plotted for both stage A (blue lines) and stage B (black lines). We find that the Posterior distribution does not significantly change after changing the Prior distributions. 

\begin{table}
\small
\centering
\caption{The mean, standard deviation, 2.5th percentile bound and 97.5th percentile bound are presented for each parameter for the best model for each phase.} 
\label{tab:final_parameters} 
\begin{tabular}{ccccc}
Parameter & Mean & Std. Dev. & 2.5\% CI & 97.5\% CI\\
\\
$\boldsymbol{\alpha}$\textbf{-phase} & & & &\\
\hline
$\theta_D$ & 206.45 & 0.28 & 205.93 & 207.02\\
$a_2$ & $1.41\times 10^{-3}$ & $7.58\times 10^{-5}$ & $1.26\times 10^{-3}$ & $1.56\times 10^{-3}$\\
$a_3$ & $9.71\times 10^{-6}$ & $2.29\times 10^{-7}$ & $9.26\times 10^{-6}$ & $1.02\times 10^{-5}$\\
$a_4$ & $-5.61\times 10^{-9}$ & $2.44\times 10^{-10}$ & $-6.07\times 10^{-9}$ & $-5.12\times 10^{-9}$\\
$a_5$ & $1.10\times 10^{-12}$ & $7.63\times 10^{-14}$ & $9.41\times 10^{-13}$ & $1.24\times 10^{-12}$\\
$\alpha_{Ade1952}$ \cite{Adenstedt1952_HfCpRoomT} & 1.111 & 0.532 & 0.348 & 2.402\\
$\alpha_{Aru1972}$ \cite{Arutyunov1972_Hf_Cp_highT} & 0.569 & 0.138 & 0.340 & 0.876\\
$\alpha_{Bur1958}$ \cite{Burk1958_HfCpLowT} & 0.188 & 0.023 & 0.147 & 0.237\\
$\alpha_{Cag2008}$ \cite{Cagran2008_Hf_H_HighT} & 0.533 & 0.217 & 0.210 & 1.046\\
$\alpha_{Cez1974}$ \cite{Cezairliyan1975_Hf_Cp} & 4.137 & 0.959 & 2.613 & 6.343\\
$\alpha_{Col1971}$ \cite{Collings1971_H_Cp_lowT} & 0.209 & 0.040 & 0.140 & 0.295\\
$\alpha_{Fie1961}$ \cite{Fieldhouse1961_Hf_H} & 0.444 & 0.118 & 0.250 & 0.709\\
$\alpha_{Fil1971}$ \cite{Filipov1971_Hf_Cp_highT} & 0.640 & 0.163 & 0.367 & 1.002\\
$\alpha_{Gol1970}$ \cite{Golutvin1970_Hf_Cp} & 0.703 & 0.152 & 0.449 & 1.043\\					
$\alpha_{Haw1963}$ \cite{Hawkins1963_Hf_H_highT} & 1.512 & 0.234 & 1.101 & 2.027\\	 	 	 	 
$\alpha_{Kats1985}$ \cite{Kats1985_Hf_H_highT} & 0.317 & 0.080 & 0.183 & 0.497\\
$\alpha_{Kne1963}$ \cite{Kneip1963_Hf_Cp_lowT} & 0.004 & 0.001 & 0.003 & 0.005\\
$\alpha_{Mil2006S1}$ \cite{Milošević2006_Hf_Cp_highT} & 2.409 & 0.427 & 1.670 & 3.351\\
$\alpha_{Mil2006S2}$ \cite{Milošević2006_Hf_Cp_highT} & 11.272 & 2.225 & 7.361 & 16.081\\
$\alpha_{Pel1971}$ \cite{Peletskii1971_Hf_Cp_highT} & 1.060 & 0.183 & 0.744 & 1.457\\
$\alpha_{McC1964}$ \cite{McClaine1964_Hf_Cp_lowT} & 0.053 & 0.008 & 0.038 & 0.071\\
$\alpha_{Wol1957}$ \cite{Wolcott1957_Hf_Cp_lowT} & 0.138 & 0.018 & 0.106 & 0.175\\
\\
$\boldsymbol{\beta}$\textbf{-phase} &&&&\\
\hline
$b_0$ & $-3.36\times 10^4$ & $4.27\times 10^3$ & $-4.17\times 10^4$ & $-2.44\times 10^4$\\
$b_1$ & 77.43 & 5.92 & 64.69 & 88.71\\
$b_2$ & $-4.59\times 10^{-2}$ & $5.48\times 10^{-3}$ & $-5.63\times 10^{-2}$ & $-3.42\times 10^{-2}$\\
$b_3$ & $1.20\times 10^{-5}$ & $1.00\times 10^{-6}$ & $9.00\times 10^{-6}$ & $1.50\times 10^{-5}$\\
$\alpha_{Cag2008}$ \cite{Cagran2008_Hf_H_HighT} & 0.717 & 0.260 & 0.315 & 1.324\\
$\alpha_{Cez1974}$ \cite{Cezairliyan1975_Hf_Cp} & 4.144 & 1.017 & 2.507 & 6.475\\
$\alpha_{Fil1971}$ \cite{Filipov1971_Hf_Cp_highT} & 0.616 & 0.472 & 0.075 & 1.834\\
$\alpha_{Kat1985}$ \cite{Kats1985_Hf_H_highT} & 0.123 & 0.040 & 0.059 & 0.215\\
$\alpha_{Mil2006S1}$ \cite{Milošević2006_Hf_Cp_highT} & 0.792 & 0.285 & 0.350 & 1.451\\
$\alpha_{Mil2006S2}$ \cite{Milošević2006_Hf_Cp_highT} & 1.090 & 0.532 & 0.349 & 2.368\\
$\alpha_{Par2003}$ \cite{Paradis2003_Hf_Cp_highT} & 6.364 & 2.249 & 2.816 & 11.411\\
$\alpha_{Pel1971}$ \cite{Peletskii1971_Hf_Cp_highT} & 1.025 & 0.343 & 0.501 & 1.846\\
$\alpha_{Ros2001}$ \cite{Rösner-Kuhn2001_Hf_H_highT} & 2.368 & 0.376 & 1.718 & 3.171\\
\\
\textbf{liquid-phase} &&&&\\
\hline
$c_0$ & $5.86\times 10^3$ & $1.76\times 10^3$ & $2.46\times 10^3$ & $9.30\times 10^3$\\
$c_1$ & 29.20 & 1.02 & 27.18 & 31.15\\
$c_2$ & $5.28\times 10^{-3}$ & $2.97\times 10^{-4}$ & $4.73\times 10^-3$ & $5.88\times 10^-3$\\
$\alpha_{Kor2005}$ \cite{Korobenko2005_Hf_Cp_highT} & 1.880 & 0.627 & 0.946 & 3.397\\
$\alpha_{Par2003}$ \cite{Paradis2003_Hf_Cp_highT} & 0.120 & 0.043 & 0.055 & 0.222\\
$\alpha_{Cag2008}$ \cite{Cagran2008_Hf_H_HighT} & 5.183 & 1.450 & 2.839 & 8.403\\
$\alpha_{Ros2001}$ \cite{Rösner-Kuhn2001_Hf_H_highT} & 0.713 & 0.126 & 0.496 & 0.991\\
\end{tabular}
\end{table}

\begin{figure}
\begin{subfigure}{0.5\textwidth}
  \centering
  \includegraphics[width=1.0\textwidth]{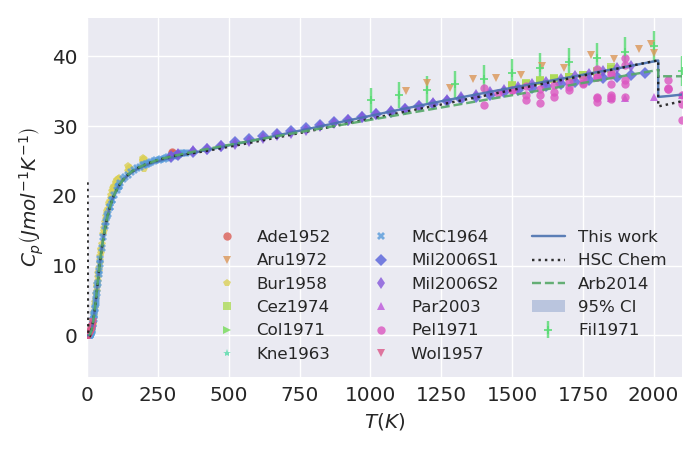}
  \caption{}\label{fig:final_alpha_Cp}
\end{subfigure}
\begin{subfigure}{0.5\textwidth}
  \centering
  \includegraphics[width=1.0\textwidth]{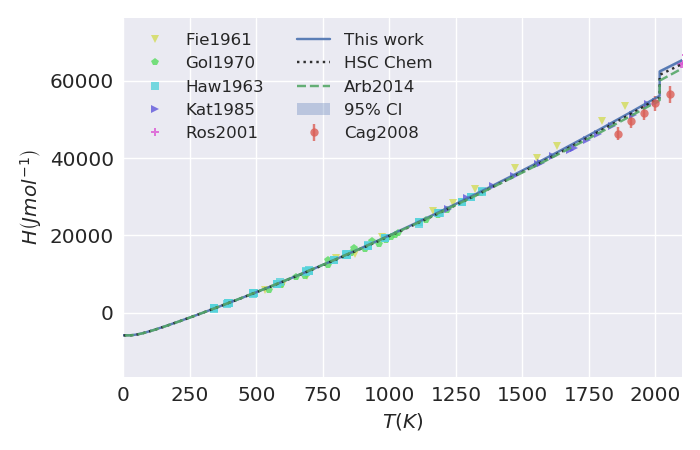}
  \caption{}\label{fig:final_alpha_H}
\end{subfigure}
\begin{subfigure}{0.5\textwidth}
  \centering
  \includegraphics[width=1.0\textwidth]{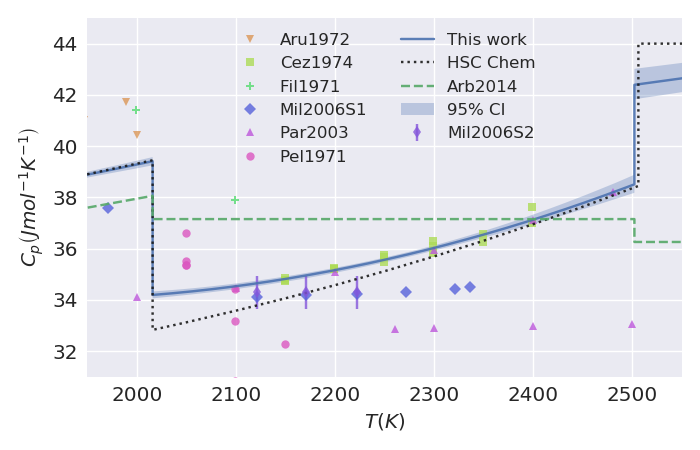}
  \caption{}\label{fig:final_beta_Cp}
\end{subfigure}
\begin{subfigure}{0.5\textwidth}
  \centering
  \includegraphics[width=1.0\textwidth]{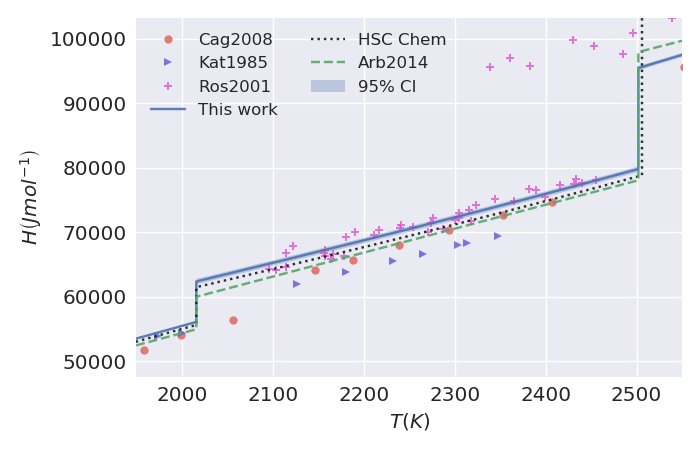}
  \caption{}\label{fig:final_beta_H}
\end{subfigure}
\begin{subfigure}{0.5\textwidth}
  \centering
  \includegraphics[width=1.0\textwidth]{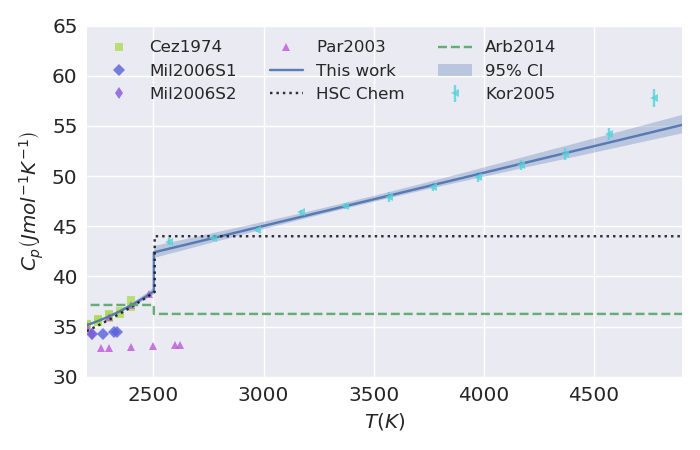}
  \caption{}\label{fig:final_liquid_Cp}
\end{subfigure}
\begin{subfigure}{0.5\textwidth}
  \centering
  \includegraphics[width=1.0\textwidth]{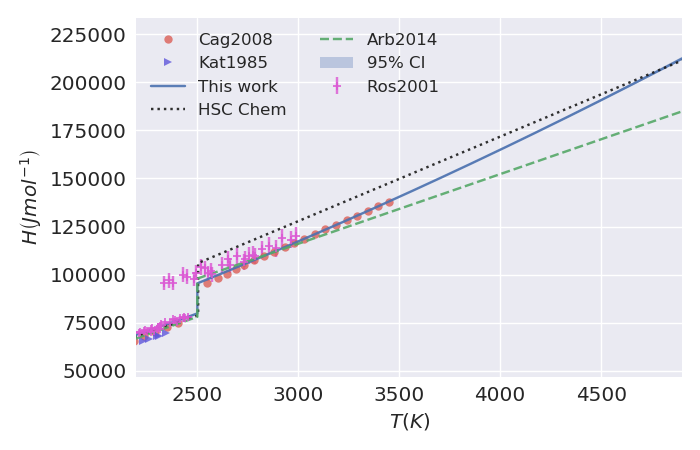}
  \caption{}\label{fig:final_liquid_H}
\end{subfigure}
\caption{The Bayesian model predictions for each of the phases versus temperature for specific heat (a, c, e) and enthalpy (b, d, f) are plotted as a solid blue line. The results of previous Hf assessments from Arb2014 and HSC Chem are plotted as dashed green lines and dotted black lines respectively. The experimental data points are plotted as markers and labeled in the legend. The error bars have been rescaled by the associated best fit hyperparameter.}
\label{fig:final_prediction}
\end{figure}

\begin{figure}
\centering
\includegraphics[width=.7\linewidth]{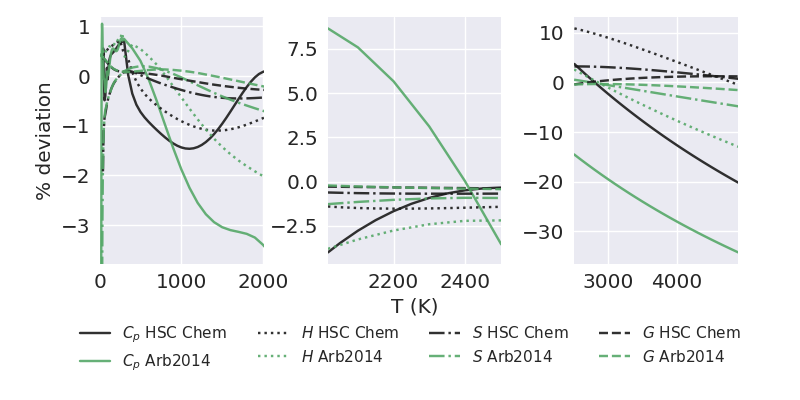}
\caption{The percentage differences between the assessments of HSC Chemistry and Arb2014 and the Bayesian model predictions given in this work are presented for each phase and thermodynamic quantity versus temperature.}
\label{fig:final_differences}
\end{figure}

Table \ref{tab:final_parameters} contains the means, standard deviations and uncertainty intervals for the parameters in Eq. (\ref{eqn:Cp_alpha}) through Eq. (\ref{eqn:H_liquid}) as well as for the hyperparameters which rescale the errors for each experimental dataset. In Fig. \ref{fig:final_prediction}, we plot the model predictions for each phase versus temperature for the results of two previous assessments and the datasets used. The first assessment is from Arblaster \cite{Arblaster2014_HfAssessment}, hereafter Arb2014, whose results leverage their knowledge and expertise to judge the quality of various experimental measurements and come up with a final model. The second assessment was performed using thermodynamic information and HSC Chemistry, a software package which automatically combines existing thermodynamic assessments to find an optimal model, leveraging multiple experts whenever possible. The software uses data from literature, especially review articles.  Whenever possible, we trace back the original article and use the original experimental measurement information. At the time of this article, HSC Chemistry does not provide uncertainty intervals for the model parameters or property model predictions. To acquire insights into the effect of the method presented in Sec. \ref{subsec:data_quality} on the actual experimental datasets, we provide error-bars for a single dataset in each subfigure. Note that these errorbars have been rescaled from the reported values by the associated best-fit hyperparameter found by our model. Percent deviations of the thermodynamic quantities between our models and the two other assessments are shown in Fig. \ref{fig:final_differences}.

Our alpha-phase model prediction for specific heat shows reasonable agreement with both the HSC Chemistry and Arb2014 assessments up until 1000K, after which Arb2014 trends consistently lower until the phase transition. The discrepancies are far smaller for enthalpy. From Fig. \ref{fig:final_alpha_Cp} and Fig. \ref{fig:final_alpha_H}, we can see that our model closely tracks the experimental measurements of Mil2006S2, Haw1963 and Cez1974. In Arb2014, the alpha-phase model is built (for 298.15K and above) solely from the measurements of Haw1963 and Kat1985. The uncertainty intervals for our alpha-phase model prediction are narrow which is likely due to the large number of experimental data points available in the alpha phase. Also note that the prediction of our model is not skewed by a number of discrepant measurements at high temperatures including Aru1972 and Cag2008.

Figures \ref{fig:final_beta_Cp} and \ref{fig:final_beta_H} make it apparent that our beta-phase model prioritizes the experimental measurements of Par2003, Cez1974 and Ros2001. This observation is corroborated by the high values of the hyperparameters for these data sets in Table \ref{tab:final_parameters}. The lower set of points of Par2003 are from measurements in the liquid phase; the temperature ranges of the different phases overlap as measurements were taken in the metastable regime. In contrast, Arb2014 selected a linear model calibrated solely on the enthalpy measurements of Ros2001 (note that the enthalpy difference in the model of Arb2014 may result from differences in the Zr content correction). This explains why the two beta-phase models follow the same trend in enthalpy but are highly discrepant in specific heat. Furthermore, the linear model of Arb2014 for enthalpy forces the specific heat to be constant and to take a value which does not pass through the majority of the experimental specific heat measurements. The difference between our model and the HSC chemistry assessment is small, as the HSC Chemistry software appears to have incorporated specific heat measurements. Uncertainty intervals are broader for the beta-phase model due to the fewer number of data points and their significant spread.

Based on Fig. \ref{fig:final_liquid_Cp} and Fig. \ref{fig:final_liquid_H}, the Bayesian framework prioritizes the measurements of Kor2005 and Cag2008 for the construction of the liquid-phase model and downweights the measurements of Par2003. As expected, these datasets have the largest and smallest hyperparameter values, respectively, in Table \ref{tab:final_parameters}. As for the beta-phase, Arb2014 selected a linear model for the liquid-phase based on the measurements from Ros2001, resulting in a constant model for specific heat which is highly discrepant from our model. HSC Chemistry also selects a linear model, albeit with a slope that is more consistent with the measurements of Kor2005 for specific heat. The origins of the discrepant enthalpy shift in the HSC Chemistry assessment is unknown, but we estimate that this is due to choices made when correcting for Zr content. The uncertainty intervals of our liquid-phase model are larger than those of the alpha- or beta-phase models due to the relative scarcity of experimental measurements.

\begin{figure}
\centering
\includegraphics[width=.6\linewidth]{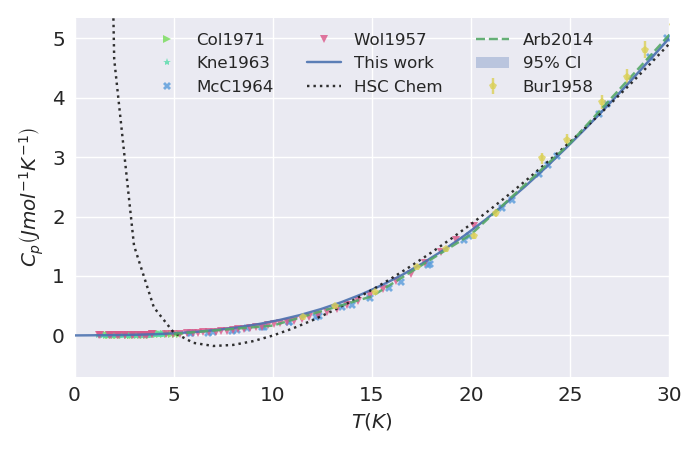}
\caption{The Bayesian model prediction (solid blue line), the results of previous Hf assessments (dashed lines) and the experimental data points (given by symbols in the legend)  versus temperature.}
\label{fig:final_alpha_Cp_lowT}
\end{figure}

Figure \ref{fig:final_alpha_Cp_lowT} shows the predicted alpha phase model at very low temperatures. The alpha-phase model does a reasonable job fitting the low temperature specific heat measurements. That said, the Debye model that dominates in this temperature regime does not have the flexibility to exactly fit the experimental data. An artifact of this discrepancy is that the hyperparameters for the low temperature datasets are close to zero. For example, the mean value of the marginalized hyperparameter for Kne1963 is 0.004, implying that errors were under-reported. Since these hyperparameter values result from model inadequacy, they do not necessarily represent errors in the experimental dataset with respect to the true thermodynamic properties. Our optimized value of the Debye temperature parameter, $\theta_D = 206.45 \pm 0.48K$, deviates significantly from what is reported by other authors (250K) \cite{Arblaster2014_HfAssessment}. This discrepancy results from employing the full Debye model across the entire temperature range for the alpha phase, whereas previous researchers determined the Debye temperature from a linearization of the specific heat versus temperature below 20K based on the low temperature limit of the Debye relationship \cite{Kneip1963_Hf_Cp_lowT}. To confirm that this is the origin of the discrepancy rather than some implicit error in the Bayesian approach, we have implemented the traditional analysis technique within our Bayesian framework and find a best fit value of $\theta_D = 253.07 \pm 0.26$ which agrees with previous results \cite{Arblaster2014_HfAssessment}. We find that the Debye temperature of the selected alpha-phase model is not dependent on the temperature range considered (i.e. our Debye temperature parameter does not significantly change when only measurements at 100K and below are included). In future work, we  plan to employ more sophisticated relationships in the low temperature region. Our low temperature model compares reasonably well with Arb2014 but shows significant discrepancies with the HSC Chemistry calculations. We attribute the differences to the negative specific heats predicted by HSC Chemistry at around 7K. 

\begin{figure}
\begin{subfigure}{0.5\textwidth}
  \centering
  \includegraphics[width=1.0\textwidth]{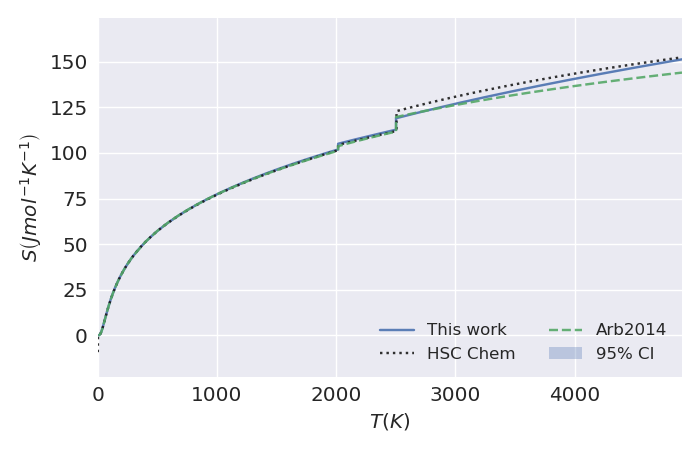}
  \caption{}\label{fig:final_all_S}
\end{subfigure}
\begin{subfigure}{0.5\textwidth}
  \centering
  \includegraphics[width=1.0\textwidth]{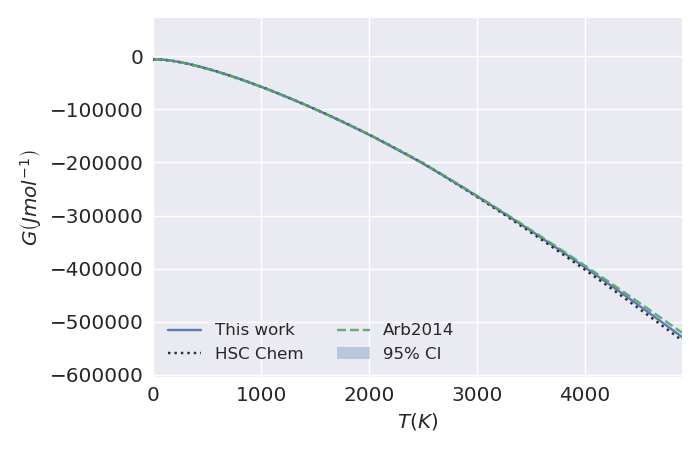}
  \caption{}\label{fig:final_all_G}
\end{subfigure}
\caption{The model predictions for all phases and the previous Hf assessments are plotted versus temperature in Kelvin for (a) entropy and (b) Gibbs free energy.}
\end{figure}

Figures \ref{fig:final_all_S} and \ref{fig:final_all_G} compare our results for entropy and Gibbs free energy for all phases against the assessments of HSC Chemistry and Arb2014. The results of the three assessments appear to be reasonably close in the alpha and beta regimes, with differences in slope and vertical placement becoming more apparent at high temperatures.

To further investigate the differences between the three assessments, we present the percentage differences between the HSC Chemistry and Arb2014 results and our model in Fig. \ref{fig:final_differences}. Each phase is plotted separately due to the significant differences in scale on the x- and y- axes. In general, errors decrease in severity for quantities in the order of specific heat, enthalpy, entropy and Gibbs free energy. This is not surprising due to the mathematical relationships between the thermodynamic quantities. Also, in most cases, the Arb2014 result is more discrepant than that of HSC Chemistry. For most of the alpha-phase temperature range, deviations are less than 4\%, however, at very low temperatures the deviation between our model and Arb2014 approaches 30\% due to the inadequacy of the Debye model as previously discussed. We do not evaluate the discrepancy for the HSC Chemistry result below 40K due to its highly anomalous behavior. Deviations between our model and the Arb2014 assessment in Cp reach 8\% for the beta-phase due to the constant behavior of the Arb2014 model. Discrepancies are even larger in the liquid-phase for both HSC Chemistry and Arb2014.

\begin{figure}
\begin{subfigure}{0.5\textwidth}
  \centering
  \includegraphics[width=1.0\textwidth]{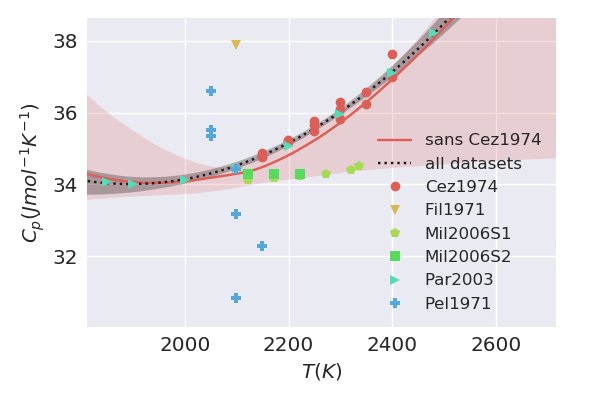}
  \caption{}\label{fig:final_beta_1out_Cp}
\end{subfigure}
\begin{subfigure}{0.5\textwidth}
  \centering
  \includegraphics[width=1.0\textwidth]{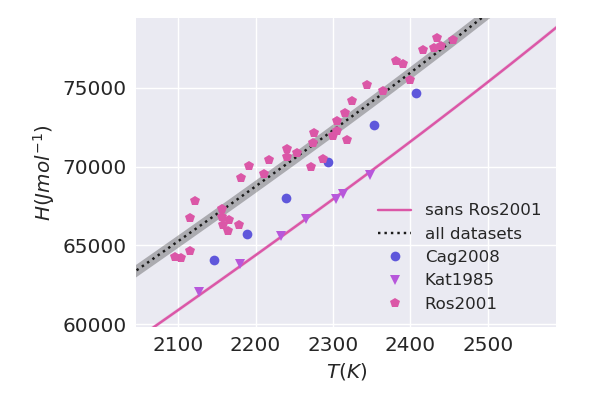}
  \caption{}\label{fig:final_beta_1out_H}
\end{subfigure}
\caption{The beta-phase models and 95\% uncertainty intervals resulting from Bayesian analyses are plotted versus temperature for (a) specific heat disregarding Cez1974, and (b) enthalpy disregarding Ros2001. The overall beta-phase prediction and 95\% uncertainty intervals are also plotted alongside the experimental measurements versus temperature.}
\label{fig:final_beta_1out}
\end{figure}

Finally, we evaluate the effect of removing individual datasets on the Bayesian analysis. For each phase, we evaluate an equal number of models as the number of relevant datasets. Each iteration, we remove a single dataset and perform the Bayesian analysis. We directly employ the model forms of Eq. (\ref{eqn:final_expressions}) in this analysis, so it is possible that different model forms would be more appropriate depending on the specific dataset dropped in each analysis. Figure \ref{fig:final_beta_1out} highlights the results of this analysis for the beta-phase. Figure \ref{fig:final_beta_1out_Cp} presents the effects of removing the measurements of Cez1974 (highly favored by the framework in the original Bayesian analysis) for specific heat. When the dataset is removed, the 50th percentile model behavior changes moderately and the 95th percentile uncertainty interval signals a drastically increased uncertainty in the model parameters. Interestingly, the Mil2006S1 and Par2003 hyperparameter distributions for this model (not presented) are bimodal, indicating that the Bayesian analysis results in comparable probabilities of these two highly-discrepant datasets having low error (i.e. being correct). Finally, Fig. \ref{fig:final_beta_1out_H} presents the effect of removing the measurements of Ros2001 on the beta-phase model for enthalpy. While the slope does not appreciably change, the Bayesian framework prioritizes the measurements of Kats1985, resulting in a significant change in the enthalpy of transformation. We observe similar changes in the alpha and liquid phase models when dropping datasets which the framework originally prioritizes. These changes highlight the importance of considering all available measurements in the assessment, and motivate additional computational analyses to help establish expected trends in material properties.

The thermodynamic assessment of Hafnium metal shows that our single framework directly accepts raw, experimental measurements of specific heat and enthalpy, and uses them to simultaneously calibrate models, produce evaluation metrics and provide uncertainty intervals for the parameters and model predictions. Furthermore, the framework can naturally account for tasks that are typically restricted to experts, such as managing outliers, rescaling inaccurate error bars, weighting datasets and finding consistency among multiple datasets expressed via different quantities. Through examination of the Posterior distributions, we can provide both model parameter uncertainties as well as model comparison using Bayes’ Factors.

\section{Summary and Conclusions}
\label{sec:summary_and_conclusions}

In this work we introduce a Bayesian framework for the selection, calibration and quantification of uncertainty of thermodynamic property models. We present the Bayesian framework in detail, and describe modifications to address common data issues in thermodynamic assessment including the presence of outliers, systematic errors, inaccurately reported errors and inconsistencies between thermodynamic quantities. Enabled by cutting-edge numerical sampling algorithms, the framework simultaneously performs these tasks in a single analysis. We accompany the framework and modifications with example problems that clearly demonstrate each method and its benefits. Perhaps most critically, the Bayesian framework provides the researcher with a clear and self-consistent method to perform assessment. Data sets, irrespective of the thermodynamic quantity measured, can be incorporated into the framework without modification, which then determines the best weights and parameters given the model. Our confidence in the best fitting parameters are described by the Posterior distributions.

We culminate the study with an assessment of the thermodynamic properties of Hafnium metal from 0 to 4900K through three distinct phases. The resulting models are broadly effective over a wide range of temperatures and come with parameter and prediction uncertainty intervals. We find that the Debye model plus a quartic polynomial performs best to model the specific heat versus temperature for the low temperature alpha phase. The model fits the data well, except at very low temperatures where the Debye model has deficiencies in modeling the thermodynamic properties. This model compares well with those of HSC Chemistry and Arb2014 through the majority of the temperature range, showing the largest percent deviations with Arb2014 at higher temperatures.  We find that  a quadratic model for specific heat for the beta phase captures the trends of the Cez1974, Par2003 and Ross2001 measurements. While this model is comparable to the HSC Chemistry prediction, larger deviations are seen versus the Arb2014 model as that work employed a linear model for enthalpy versus temperature and therefore a constant model for specific heat. Our results indicate that a linear model performs well at describing the temperature dependence of specific heat in the liquid phase, aligning most closely with the Kor2004 and Cag2008 data sets. Our liquid phase model shows significant discrepancies with HSC Chemistry and Arb2014 for specific heat as both employ linear models to describe the dependence of enthalpy versus temperature, and therefore have constant specific heat versus temperature. These discrepancies highlight the Bayesian framework's ability to select and calibrate models, optimally leveraging both specific heat and enthalpy measurements. The approaches presented in this work have direct application to future efforts to improve the assessment of binary, ternary and multi-component systems, lessening the burden on the expert in managing unwieldy collections of data. 

\section*{Acknowledgments}

The authors thank the CHiMaD Program funded by the National Institute for Standards and Technology
(NIST) for supporting this study. We also thank the Laboratory Computing Resource Center at Argonne
National Laboratory for providing computational time and assistance running software on the Bebop compute
cluster.

\section*{Appendix A}
\label{sec:appendix_a}

In this section, we demonstrate Bayesian model selection, calibration and uncertainty quantification through a simple example problem. We employ synthetic data because it provides a convenient ground truth for comparison. In this example, we generate 20 pairs of x and y values according to the following relationships,
\begin{subequations} \label{eqn:ex1_true_model}
\begin{gather}
x\sim\mathcal{U}\left(\left[0,1\right]\right),   \\
\varepsilon\sim\mathcal{N}\left(0,0.05\right),   \\
y=1+x+\varepsilon,
\end{gather}
\end{subequations}
where $\mathcal{U}$ is the uniform distribution between [0, 1) and $\mathcal{N}$ is the Normal (Gaussian) distribution with mean 0 and variance 0.05.

In the first step of the Bayesian analysis we select a set of potential models. One of the strengths of this framework is that models of any form with different numbers of parameters may be compared. In this example, we select polynomials of order $N$ represented as follows,
\begin{equation} \label{eqn:ex1_model}
M\left(x,\boldsymbol{\Theta}\right)=\sum_{i=0}^N\Theta_i x^i,
\end{equation}
where $i$ is the exponent of $x$ and indexes the model parameter vector. Next, we define the Likelihood and priors on the model parameters. We assume independence of the data points and write the Likelihood as the product of the individual Gaussian Likelihoods for each data point,
\begin{equation} \label{eqn:ex1_likelihood}
P\left(\boldsymbol{D}|\boldsymbol{\Theta},\varepsilon,M\right)=\prod_j\mathcal{N}\left(y_j|M\left(x_j,\boldsymbol{\Theta}\right),\varepsilon\right),
\end{equation}
where $i$ indexes each data point and $\varepsilon$ is standard error parameter, which is included in the Bayesian analysis and captures the standard deviation of the data generation process. Note that, in practice, the product is replaced by a summation of log-Likelihoods; this alleviates the issues of arithmetic underflow and overflow that are common when multiplying large sets of probability densities. Priors are chosen based on the best available knowledge about the parameters and the relationships between parameters. In this simple example we select uniform priors as follows,
\begin{subequations} \label{eqn:ex1_prior}
\begin{gather}
P\left(\Theta_i|M\right)=\mathcal{U}\left(\Theta_i|-2,2\right),   \\
P\left(\varepsilon|M\right)=\mathcal{U}\left(\varepsilon|0,1\right).
\end{gather}
\end{subequations}

\begin{figure}
\begin{subfigure}{0.5\textwidth}
  \centering
  \includegraphics[width=1.0\textwidth]{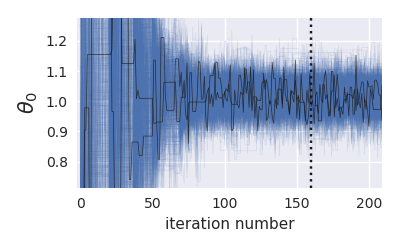}
  \caption{}\label{fig:ex1_chains}
\end{subfigure}
\begin{subfigure}{0.5\textwidth}
  \centering
  \includegraphics[width=1.0\textwidth]{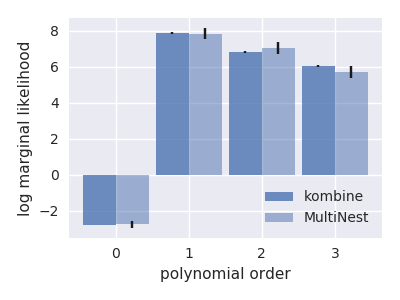}
  \caption{}\label{fig:ex1_evidence}
\end{subfigure}
\caption{(a) Sequences of 400 walkers from the kombine sampler over 150 iterations. Two randomly selected sequences are highlighted in black. All samples to the left of the black dashed line are discarded. (b) The marginal Likelihoods and 2-sigma standard deviations are plotted for both samplers and all models.}
\label{fig:ex1_samplers}
\end{figure}

By defining the Likelihood and priors we have specified the Bayesian model. Even though there are different numerical techniques available to evaluate the Posterior, the result should not depend on the sampler. As a practical demonstration, we employ both samplers to evaluate the posterior distributions and marginal Likelihoods for polynomial models of order 0 through 3. First, we initialize 400 points (denoted walkers and active points for kombine and MultiNest, respectively) from the prior distributions. We monitor the kombine sequences for convergence based on the GR metric introduced Sec \ref{subsec:numerical_sampling}. Fig \ref{fig:ex1_chains} shows the paths of the walkers in kombine for the first parameter in the 1st order polynomial model. Note that the sampler automatically discards all samples to the left of the black dashed line to allow the walkers to begin randomly exploring the Posterior area (i.e. burn-in). Figure \ref{fig:ex1_evidence} plots the marginal Likelihoods and $2\sigma$ standard deviation bounds for each sampler and polynomial model. Based on the computed Bayes Factors, the Evidence for the linear model is weak versus the quadratic, positive versus the cubic, and very strong versus the constant model. Consequently, it would be reasonable to select the linear model as it exhibits the highest marginal Likelihood and has fewer parameters than the next best (the quadratic) model. Furthermore, the marginal Likelihoods agree within the errors for both samplers.

\begin{figure}
\begin{subfigure}{0.5\textwidth}
  \centering
  \includegraphics[width=1.0\textwidth]{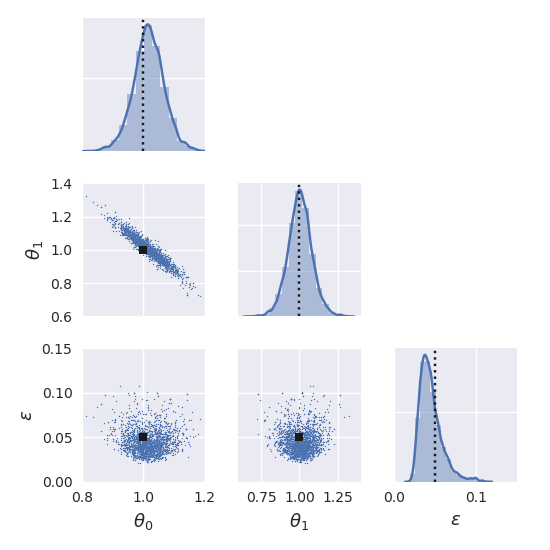}
  \caption{}\label{}
\end{subfigure}
\begin{subfigure}{0.5\textwidth}
  \centering
  \includegraphics[width=1.0\textwidth]{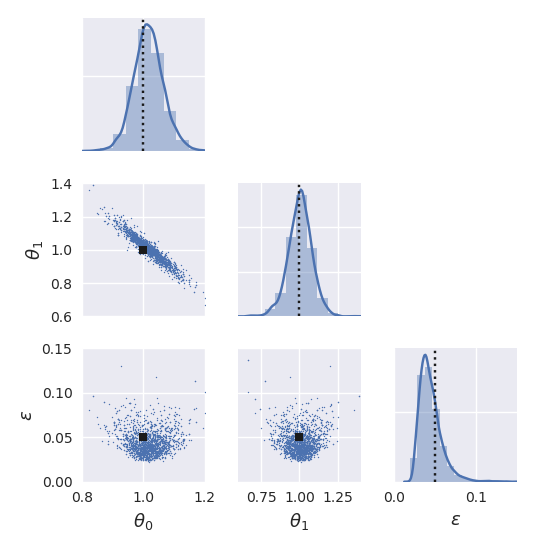}
  \caption{}\label{}
\end{subfigure}
\caption{Histograms, kernel density estimated distributions and scatter plots of the posterior distributions obtained via (a) kombine, and (b) MultiNest, respectively. The true parameter values are marked as a black dashed line.}
\label{fig:ex1_posterior_param}
\end{figure}

\begin{figure}
\begin{subfigure}{0.5\textwidth}
  \centering
  \includegraphics[width=1.0\textwidth]{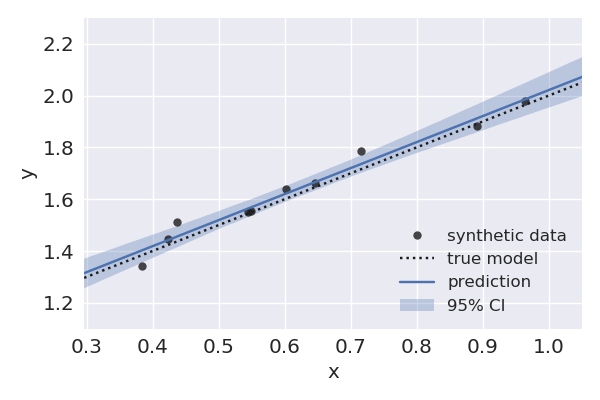}
  \caption{}\label{}
\end{subfigure}
\begin{subfigure}{0.5\textwidth}
  \centering
  \includegraphics[width=1.0\textwidth]{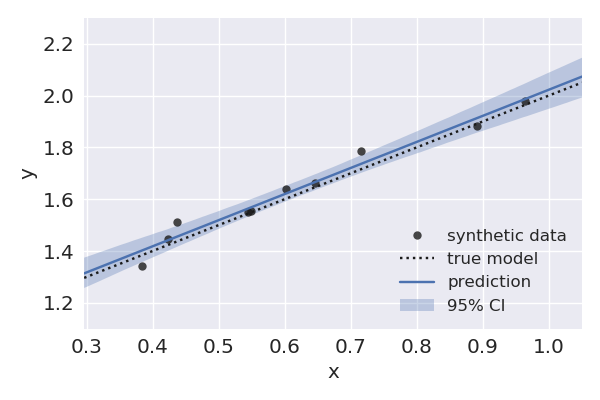}
  \caption{}\label{}
\end{subfigure}
\caption{The model prediction, uncertainty intervals, true model and synthetic data points are plotted versus x for (a) kombine and (b) MultiNest, respectively.}
\label{fig:ex1_posterior_predict}
\end{figure}

Finally, we examine the Posterior distribution and uncertainty intervals. Figure \ref{fig:ex1_posterior_param} summarizes the posterior distributions obtained via the kombine (left) and MultiNest (right) samplers. We plot histograms and KDE distributions for each univariate parameter distribution. We provide multivariate summaries for each pair of coefficients via scatter plots of 2000 randomly selected samples. True parameter values are indicated by black markers. We first note that the samplers produce similar posterior parameter distributions. Furthermore, regions of high density in the Posterior closely align with the true parameter values. Figure \ref{fig:ex1_posterior_predict} displays the model prediction and 95$^{\rm th}$ percentile uncertainty intervals obtained via the kombine (left) and MultiNest (right) samplers versus the synthetic data points. The uncertainty intervals are obtained by randomly sampling the Posterior distributions of the model parameters and computing the 2.5$^{\rm th}$ and 97.5$^{\rm th}$ percentile levels of the resulting model predictions. The uncertainty intervals show the expected spread of model predictions from the distribution of $\theta_0$ and $\theta_1$. The models fit the synthetic data well, and both the model predictions and uncertainty intervals are nearly indistinguishable between the two samplers. The results of Fig. \ref{fig:ex1_evidence}, Fig. \ref{fig:ex1_posterior_param}, and Fig. \ref{fig:ex1_posterior_predict} give us confidence that different samplers obtain the same posterior parameter distributions given a single set of Likelihood and prior definitions.

\section*{Appendix B}
\label{sec:appendix_b}

\begin{figure}
\begin{subfigure}{0.5\textwidth}
  \centering
  \includegraphics[width=1.0\textwidth]{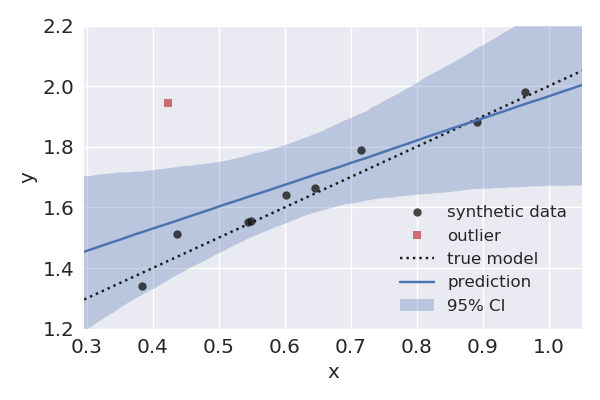}
  \caption{}\label{fig:ex2_normL_pred}
\end{subfigure}
\begin{subfigure}{0.5\textwidth}
  \centering
  \includegraphics[width=1.0\textwidth]{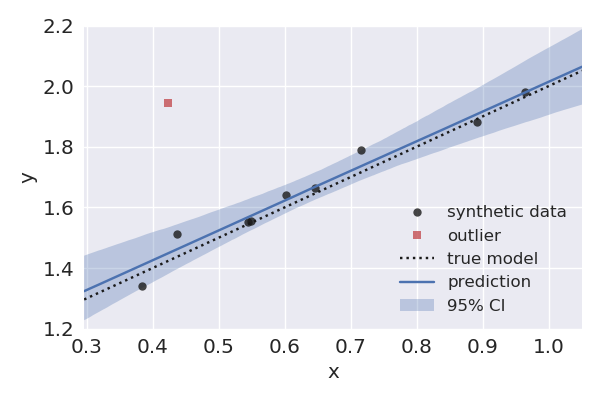}
  \caption{}\label{fig:ex2_t2L_pred}
\end{subfigure}
\caption{The model prediction, uncertainty interval, true model and synthetic data points are plotted with (a) Normal Likelihood (b) Student-t Likelihood with $2 + 10^{-6}$ degrees of freedom.}
\label{fig:ex2}
\end{figure}

In this section we demonstrate the ability of the Bayesian framework to accommodate outliers through the analysis of a data set obtained by perturbing a single datum from Appendix A, shown as a red square in Fig. \ref{fig:ex2}. Fig. \ref{fig:ex2_normL_pred} shows the results of this analysis using the Likelihood given in Eq. (\ref{eqn:ex1_likelihood}). Clearly, the model prediction is biased by the outlier, and the uncertainty intervals show significant spreads in the slope and intercept. In this example, we replace the Normal distribution with a Student’s-t distribution and $2 + 10^{-6}$ for the degrees of freedom.

Clearly, by employing Student’s-t distribution for the Likelihood definition the robustness to the outlier is greatly improved. The model prediction is comparable to that from Fig. \ref{fig:ex1_posterior_predict} of the previous section.

\section*{Appendix C}
\label{sec:appendix_c}

In this Section, we demonstrate the approach described in Sec. \ref{subsec:data_quality} to rescale the errors of datasets and accommodate systematic errors. We use three synthetic data sets of 10 points each. We generate Set A via Eq. (\ref{eqn:ex1_true_model}) and assign error bars equal to the standard deviation of the $\varepsilon$ standard error parameter (equal to 0.05). In this example we will assume that the errors for Set A are accurate. The points in Set B follow the same trend, but with a 0.2 standard error (4 times larger than for Set A). The reported errors in Set B underestimate the standard error by a factor of 4. We generate Set C with an intercept 0.5 less than the ‘true’ value and a 0.05 standard error (also reported as 0.05). We then perform the analysis with and without the use of hyperparameters. In the method without hyperparameters, we directly use the reported standard errors in the Likelihood definition (we do not need to employ a parameter for the data variance as in Eq. (\ref{eqn:ex1_likelihood})). In the hyperparameter method, we initialize one hyperparameter for each of the three data sets.

\begin{figure}
\begin{subfigure}{0.5\textwidth}
  \centering
  \includegraphics[width=1.0\textwidth]{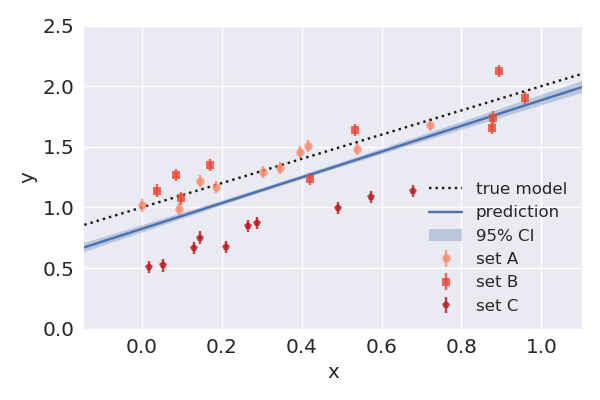}
  \caption{}\label{fig:ex3_standard_pred}
\end{subfigure}
\begin{subfigure}{0.5\textwidth}
  \centering
  \includegraphics[width=1.0\textwidth]{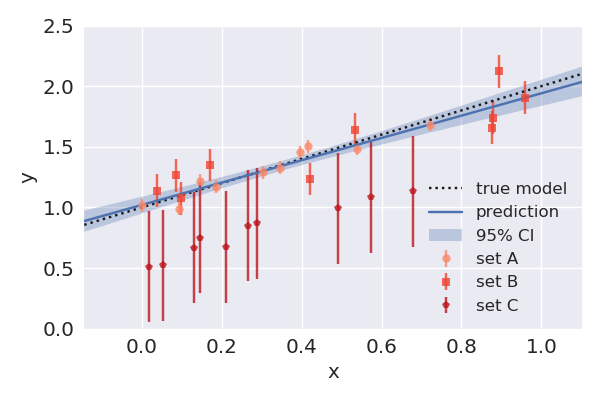}
  \caption{}\label{fig:ex3_yma_pred}
\end{subfigure}
\caption{The model prediction, uncertainty intervals and synthetic data points are plotted versus x for the posterior distributions (a) employing the reported errors, and (b) weighing the errors using hyperparameters. Note that for a visual effect we have rescaled the errorbars in (b) by the best fit hyperparameters.}
\label{fig:ex3_errorbar_pred}
\end{figure}

\begin{figure}
\centering
\includegraphics[width=.6\linewidth]{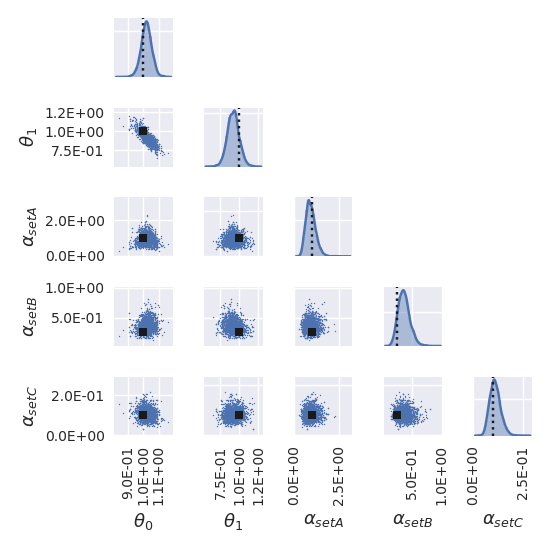}
\caption{The 2D marginal Posterior distributions obtained via the hyperparameter method. True parameter values are shown in a black dashed line.}
\label{fig:ex3_yma_cov}
\end{figure}

Figures \ref{fig:ex3_standard_pred} and \ref{fig:ex3_yma_pred} display the results of these analyses with and without hyperparameters. Without hyperparameters, the best fit model lies between data sets A and B, which follow the same trend, and set C, which has a different intercept. In contrast, the best fit model using hyperparameters, shown in Fig. \ref{fig:ex3_yma_pred} agrees with the true model used to generate the data in this example. In this figure we have scaled the errors on each dataset by the best fit hyperparameter values of 0.9, 0.4 and 0.1 for hyperparameters $\alpha_A$, $\alpha_B$ and $\alpha_C$, respectively. Figure \ref{fig:ex3_yma_cov} displays the histograms and KDE distributions for each parameter and scatter plots for each pair of parameters. Each ‘true parameter value’ (marked by the black dashed line) falls within the high probability regions of the associated Posterior distribution.  Considering the log marginal Likelihood for both models we find a values $-2.6$ for the model with hyperparameters and $-323.2$ without. In this case the Bayes’ Factor favors the model with hyperparameters.

\begin{spacing}{0.5}
\bibliography{refs}
\end{spacing}

\end{document}